\documentclass[preprint,tightenlines,showkeys,showpacs,prd,byrevtex]{revtex4} 
\usepackage{graphicx}
\usepackage{bm}
\begin{document}      
\preprint{YITP-09-03}
\title{$\Lambda(1405)$ photoprodcution with $p$-wave contributions}      
%-------------------------------------------------
\author{Seung-il Nam}
\email[E-mail: ]{sinam@yukawa.kyoto-u.ac.jp}
\affiliation{Yukawa Institute for Theoretical Physics (YITP), Kyoto
University, Kyoto 606-8502, Japan} 
%-------------------------------------------------
\author{Daisuke Jido}
\email[E-mail: ]{jido@yukawa.kyoto-u.ac.jp}
\affiliation{Yukawa Institute for Theoretical Physics (YITP), Kyoto
University, Kyoto 606-8502, Japan} 
%-------------------------------------------------
\date{\today}
\begin{abstract}  
In this report, we present our recent studies on the $\Lambda(1405,1/2^-)\equiv\Lambda^*$ photoproduction via the $\gamma p\to K^+\pi^\pm\Sigma^\mp$ scattering process, employing a coupled-channel formalism, i.e., the chiral unitary model, which respects chiral dynamics in terms of the Weinberg-Tomozawa meson-baryon interaction. In addition to the dynamically generated $s$-wave resonance, $\Lambda^*$, we include the $p$-wave resonance contribution from $\Sigma(1385,3/2^+)$. We observe that the $p$-wave contribution provides considerable effects on the $\pi\Sigma$ invariant-mass spectrum. It also turns out that, within the present level of calculations, we can reproduce hardly the recent $\Lambda^*$-photoproduction experiment by LEPS collaboration at SPring-8. 
\end{abstract} 
\maketitle
%--------------------------------------------------
\section{Introduction}
%--------------------------------------------------
To date, $\Lambda(1405,1/2^-)\equiv\Lambda^*$ has been a center of discussions on the generation of resonances with strangeness. Especially, we have had many arguments on the internal structure of $\Lambda^*$, which differs from usual color-singlet three-quark baryons for instance~\cite{Dalitz:1960du,Akaishi:2002bg}. In terms of chiral dynamics, $\Lambda^*$ is also proposed to be a quasi $\bar{K}N$ bound state as an $s$-wave resonance~\cite{Hyodo:2008xr}. Related to this interesting subject, there have been many theoretical works done in Refs.~\cite{Sekihara:2008qk,Hyodo:2007np,Jido:2008kp,KanadaEn'yo:2008wm,Williams:1991tw,Nam:2008jy,Nam:2003ch} and references therein. Experimentally, the hadron-induced productions of $\Lambda^*$ were explored in Refs.~\cite{Thomas:1973uh,Hepp:1976ky,Hemingway:1984pz,Prakhov:2004an}, whereas the photon-induced experiments have been performed mainly by LEPS collaboration at SPring-8 in Japan~\cite{Ahn:2003mv,Ahn1,Niiyama:2008rt}. 

The recent experiment for the $\Lambda^*$ photoproduction via the $\gamma p\to K^+\pi^\pm\Sigma^\mp$ scattering process, conducted by LEPS collaboration~\cite{Niiyama:2008rt}, showed considerable angular dependence of its production rate. In comparison to the previous experiment~\cite{Ahn:2003mv}, in which the outgoing pion and kaon were detected almost in a forward angle region, the new experiment measured the pion for a {\it wider} angle using the time-projection chamber (TPC), whereas the kaon was detected for a forward angle in a similar way to that of the previous one. Typical features observed in the new experiment are given as follows:
\begin{itemize}
\item There is a larger $\Lambda^*$-production rate for the $\pi^-\Sigma^+$ channel than that for the $\pi^+\Sigma^-$ one, while their production rate were similar to each other in the previous experiment. We note that the production rate for the $\pi^-\Sigma^+$ channel is almost twice in comparison to that for the $\pi^+\Sigma^-$ one in the $K^+X$ missing mass spectrum.
\item As the energy of the incident photon increases, the $\Lambda^*$-production rate decreases drastically. In other words, $\Lambda^*$ is produced mainly in the vicinity of the threshold. Considering the $\Sigma(1385,3/2^+)\equiv\Sigma^*$-photoproduction experiment by CLAS collaboration at Jafferson Lab~\cite{Guo:2006kt}, the $\Lambda^*$-production rate is about seven times larger for the region $E_\gamma\lesssim2.0$ GeV than that beyond it.
\end{itemize}   

If this is the case, the difference between the two experiments~\cite{Ahn:2003mv,Niiyama:2008rt} may tell us that the angular dependence of the $\Lambda^*$-production mechanism is of great importance. In the present work, we want to investigate this interesting observation theoretically, employing a coupled-channel formalism, i.e., the chiral unitary model ($\chi$UM), which has explained the nature of $\Lambda^*$ successfully as a $s$-wave  resonance as well as a quasi-bound state of $\bar{K}N$~\cite{Oset:1997it,Kaiser:1995eg}.  It is worth mentioning that $\Lambda^*$ has been also interpreted as a two-pole  structure within the framework~\cite{Jido:2003cb}. The $\Lambda^*$ photoproduction via the $\gamma p\to K^+\pi^\pm\Sigma^\mp$ scattering process was already explored using the $\chi$UM with only the $s$-wave meson-baryon interaction, such as the Weinberg-Tomozawa one~\cite{Nacher:1998mi}. In order to take into account the angular-dependent meson-baryon interaction, we include a $p$-wave contribution from $\Sigma(1385,3/2^+)$ additionally, as done by one of the authors (D.J.) in Ref.~\cite{Jido:2002zk}. 

Incorporating these previous works~\cite{Nacher:1998mi,Jido:2002zk}, we want to provide theoretical interpretations on the interesting recent experimental data. As numerical results, we present the  $\pi\Sigma$ invariant-mass spectrum for the two different isospin channels as well as the Dalitz plot as a function of the two invariant masses, $M_{\pi\Sigma}$ and $M_{K\pi}$. From these results, we observe that the effects from the $\Sigma^*$-pole contribution is compatible in comparison with that of $\Lambda^*$. Unfortunately, the large difference between the isospin channels is not reproduced within the present level of calculations, in which the production mechanism of $\Lambda^*$ and $\Sigma^*$ is partially considered. Moreover, the typical energy dependence is also hard to be explained. From these observations, we conclude that it is necessary to include more realistic contributions, which reflect the structural information of $\Lambda^*$, such as a diagram with a meson-baryon loop, to which the incident photon couples, etc.. 

We organize this report as follows: in Section~II we provide formalism for the $\chi$UM with the $s$- and $p$-wave contributions. The $\Lambda^*$-photoprodution amplitude is computed in the present approach in Sectoin~III. Numerical results and discussions are given in Section~IV. The final section is devoted to summary and conclusion. 
 
%--------------------------------------------------
\section{Chiral unitary model ($\chi$UM) with $p$-wave contribution}
%--------------------------------------------------
%FIGURE>>>
\begin{figure}[t]
\includegraphics[width=12cm]{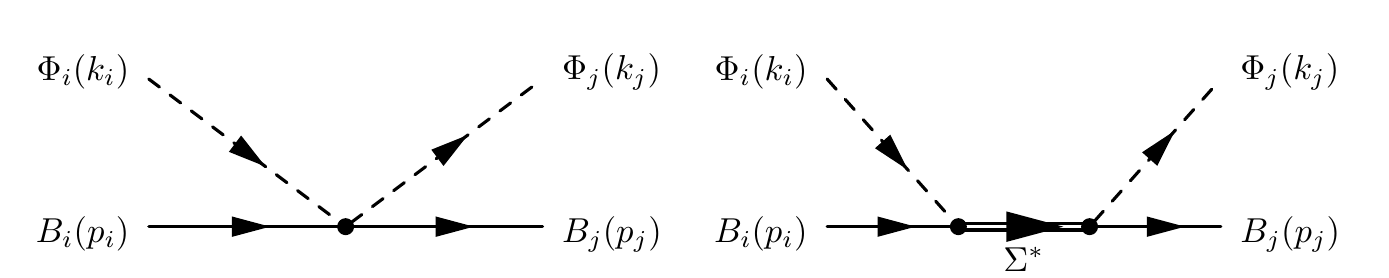}
\caption{Feynman diagrams for the meson-baryon scattering ($\Phi_{i}B_{i}\to\Phi_{j}B_{j}$) with the Weinberg-Tomozawa (left) and the $\Sigma^{*}$-pole (right) contributions. The solid and dashed lines denote the baryon and pseudoscalar meson, respectively, whereas the double-solid line the $\Sigma^{*}$.}        
\label{fig0}
\end{figure}
%FIGURE<<< 

In this section, we want to make a brief introduction for the chiral unitary model ($\chi$UM), which is a coupled-channel method based on the low-energy chiral interaction and unitarity. By virtue of the unitarity, one can explore the region beyond the ground-state baryons. Within the framework, the meson-baryon interaction can be expressed by the $s$-wave chiral Lagrangian, i.e., the Weinberg-Tomozawa (WT) interaction: 
%EQUATION>>>
\begin{equation}
\label{eq:WT}
\mathcal{L}_{\mathrm{WT}}=-\frac{iC_{ij}}{4f^2}
\bar{B}_j[\Phi_j(\rlap{/}{\partial}\Phi_i)-\Phi_i(\rlap{/}{\partial}\Phi_j)]B_i
+\mathrm{h.c.},
\end{equation}
%EQUATION<<<
where the subscripts $i$ and $j$ stand for the initial and final meson-baryon channels, respectively. $C_{ij}$ stands for a $10\times10$ coefficient matrix for SU(3) octet meson-baryon channels as given in Ref.~\cite{Oset:1997it}, whereas $B$ and $\Phi$ the $3\times3$ ground-state SU(3) octet baryon and meson fields, respectively. $f$ stands for a normalization constant of the meson fields, which is chosen to be $f=f_{\pi}\times 1.123$~\cite{Jido:2002zk}.

As mentioned previously, we will consider the $p$-wave contribution by introducing the contribution from $\Sigma(1385,3/2^+)\equiv\Sigma^*$ explicitly as done in Ref.~\cite{Jido:2002zk}. For this purpose, we make use of the following Yukawa interaction for the $\Phi B\Sigma^*$ vertex:
%EQUATION>>>
\begin{equation}
\label{eq:SIG}
\mathcal{L}_{\Phi_i B_i\Sigma^*}=
D_{i}\bar{\Sigma}^{*\mu}(\partial_{\mu}\Phi)B+\mathrm{h.c.},
\end{equation}
%EQUATION<<<
where $D_i$ denotes a diagonal matrix for the $\Phi B\Sigma^*$-coupling strength, computed from the flavor SU(3) symmetry, and can be defined as
%EQUATION>>>
\begin{equation}
\label{eq:DD}
D_{i}=\frac{12\,C^{*}_i}{5}\frac{D+F}{2f},
\end{equation}
%EQUATION<<<   
where we have used $D=0.85$ and $F=0.52$, and the $10\times10$ diagonal matrix $C^{*}$ is given in Ref.~\cite{Jido:2002zk}. Using the effective Lagrangians given in Eqs.~(\ref{eq:WT}) and (\ref{eq:SIG}), one can write an elementary meson-baryon scattering amplitudes at tree level, which are the WT and $\Sigma^{*}$-pole contributions, as depicted in Fig.~\ref{fig0}, respectively:
%EQUATION>>>
\begin{eqnarray}
\label{eq:V1}
V_{ji}&=&V_{\mathrm{WT},ji}+
[2(\hat{\bm k}_j\cdot\hat{\bm k}_i)
-i{\bm\sigma}\cdot(\hat{\bm k}_j\times\hat{\bm k}_i)]V_{\Sigma^*,ji},
\end{eqnarray}
%EQUATION<<<
where the two scalar potentials read:
%EQUATION>>>
\begin{equation}
\label{eq:V2}
V_{\mathrm{WT},ji}=-\frac{C_{ji}}{4f^2}\left(2\sqrt{s}-M_j-M_i\right),
\,\,\,\,
V_{\Sigma^*,ji}=-\frac{D_jD_i}{3}\frac{|{\bm k}_j||{\bm k}_i|}
{\sqrt{s}-E_{\Sigma^*}},
\end{equation}
%EQUATION<<<
where $E_{\Sigma^*}=(\bm{p}^2_{\Sigma^*}+M^2_{\Sigma^*})^{1/2}$. $M_{i}$ and $\sqrt{s}$ stand for the mass of the baryon in the $i$th-channel and the total energy in the center of mass frame, resepctively. Note that, in deriving Eqs.~(\ref{eq:V1}) and (\ref{eq:V2}), we have made non-relativistic reduction of the scattering amplitudes, since we are interested in the vicinity of the threshold of each channel, and taken into account only the leading contributions, which are functions of the meson momenta only. $\hat{\bm k}$ indicates the unit vector in the direction of the meson three momentum ${\bm k}$. The absolute value of the three momentum and energy for the meson in the $i$-channel can be written as:
%EQUATION>>>
\begin{equation}
\label{eq:KKKK}
|{\bm k}_i|=\sqrt{E^2_i-m^2_i},\,\,\,\,E_i=\frac{s+m^2_i-M^2_i}{2\sqrt{s}}.
\end{equation}
%EQUATION<<<
Here, $m_{i}$ stands for the mass of the meson for the $i$th-channel. Collecting all ingredients discussed so far, we can write the total scattering amplitude for $\Phi_i B_i\to\Phi_j B_j$ as follows:
%EQUATION>>>
\begin{eqnarray}
\label{eq:V}
T_{ji}&=&T_{\mathrm{WT},ji}+
[2(\hat{\bm k}_j\cdot\hat{\bm k}_i)
-i{\bm\sigma}\cdot(\hat{\bm k}_j\times\hat{\bm k}_i)]T_{\Sigma^*,ji},
\end{eqnarray}
%EQUATION<<<
where we have performed angular integrals over $\hat{\bm k}_l$, and the unitarized amplitudes for each contribution, $T_\mathrm{WT}$ and $T_{\Sigma^*}$, become:
%EQUATION>>>
\begin{equation}
\label{eq:WTSIG}
T_\mathrm{WT}=\frac{1}{[1-V_\mathrm{WT}G]}V_\mathrm{WT},\,\,\,\,
T_{\Sigma^*}=\frac{1}{[1-V_{\Sigma^*}G]}V_{\Sigma^*}.
\end{equation}
%EQUATION<<<
We note  that $G_l$ indicates the on-shell factorized meson-baryon propagator, defined as:
%EQUATION>>>
\begin{equation}
\label{eq:GF}
G_l(\sqrt{s})=i\int\frac{d^{4}q}{{(2\pi)}^{4}}
\frac{2M_l}{[(\sqrt{s}-q)^{2}-M_l^{2}](q^{2}-m_l^{2})}.
\end{equation}
%EQUATION<<<
Applying the dimensional regularization scheme to tame the loop divergence, we have the following analytic expression for $G_l$:
%EQUATION>>>
\begin{equation}
\label{eq:GDIM}
G_l(\sqrt{s})=\frac{2M_l}{16\pi^2}
\left[\frac{m^{2}_l-M^2_l+s}{2s}
\ln\frac{m^2_l}{s}+\frac{\xi_l}{2s}
\ln\frac{M^2_l+m^2_l-s-\xi_l}
{M^2_l+m^2_l-s+\xi_l}\right]+
\frac{2M_l}{16\pi^2}\ln\frac{M^2_l}{\mu^2},  
\end{equation}
%EQUATION<<<
where $\xi_l=\left[(M^2_l-m^2_l-s)^{2}-4\,s\,m^2_l\right]^{1/2}$, and $\mu$ denotes the renormalization scale, corresponding to subtraction constants in the $N/D$ method~\cite{Oset:1997it}. Converting the last term in the {\it r.h.s.} of Eq.~(\ref{eq:GDIM}) to the subtraction parameters, we have them as listed in Table.~\ref{table0}~\cite{Jido:2002zk}. We note that these parameters are determined to reproduce the $S=-1$ meson-baryon scattering data and slightly different from those given in Refs.~\cite{Oset:1997it}, since these were obtained considering the $p$-wave contributions additionally. However, the difference between the two parameter sets are almost negligible.

%TABLE>>>
\begin{table}[b]
\begin{tabular}{c|c|c|c|c|c}
$a_{\bar{K}N}$&$a_{\pi\Sigma}$&$a_{\pi\Lambda}$&$a_{\eta\Lambda}$&
$a_{\eta\Sigma}$&$a_{K\Xi}$\\
\hline
$-1.84$&$-2.00$&$-1.83$&$-2.25$&$-2.38$&$-2.67$\\
\end{tabular}
\caption{Subtraction parameters for each channel.}
\label{table0}
\end{table}
%TABLE<<<

%--------------------------------------------------
\section{Scattering amplitude for $\gamma p\to K^+\pi^\pm\Sigma^\mp$ via $\chi$UM}
%--------------------------------------------------
%FIGURE>>>
\begin{figure}[t]
\includegraphics[width=15cm]{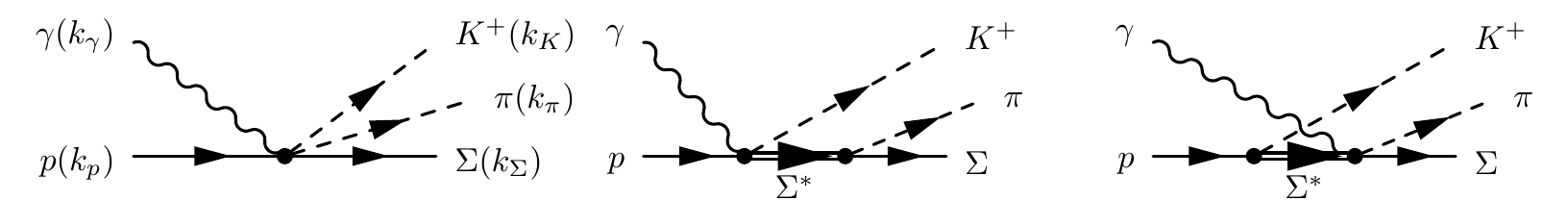}
\caption{Relevant Feynman diagrams for the $\gamma p\to K^+\pi^\pm\Sigma^\mp$ reaction process at tree level. Each diagram corresponds to the contributions, a (left), b (middle), and c (right), respectively, given in Eq.~(\ref{eq:Vg}). The line styles are the same with those shown in Fig.~\ref{fig0}.}       
\label{fig1}
\end{figure}
%FIGURE<<< 

Now, we are in a position to consider the photoprodcution of the $\Lambda(1405)$ from the  $\gamma p\to K^+\pi^\pm\Sigma^\mp$ reaction process. In order to conserve the Ward-Takahashi identity for the process, it is necessary to include all topologically possible diagrams, in which the incident photon can couple to a certain vertex. However, we take into account only the contact-interaction diagrams as shown in Fig.~\ref{fig2}, which are simply obtained by gauging the interactions of Eqs.~(\ref{eq:WT}) and (\ref{eq:SIG}). According to this simplification, the tree-level photoproduction amplitude reads in the center of mass frame:
%EQUATION>>>
\begin{eqnarray}
\label{eq:Vg}
V^{s,p}_\gamma&=&\underbrace{
i[{\bm\sigma}\cdot(\hat{\bm\epsilon}
\times\hat{\bm k}_\gamma)]V_a}_\mathrm{(a)}
\nonumber\\
&+&\underbrace{[2(\hat{\bm k}_\pi\cdot\hat{\bm\epsilon})
-i{\bm\sigma}\cdot(\hat{\bm k}_\pi\times\hat{\bm\epsilon})]
V_b}_\mathrm{(b)}
+\underbrace{[2(\hat{\bm\epsilon}\cdot\hat{\bm k}_K)
-i{\bm\sigma}\cdot(\hat{\bm\epsilon}\times\hat{\bm k}_K)]
V_c}_\mathrm{(c)}.
\end{eqnarray}
%EQUATION<<<
where the amplitudes, $V^{a,b,c}$, are written by:
%EQUATION>>>
\begin{eqnarray}
\label{eq:EVg}
V_a&=&
\frac{(e_\pi-e_K)C_{\bar{K}p\to\pi\Sigma}
|{\bm k}_{\gamma}-{\bm k}_K|}{8f^2M_N},
\nonumber\\
V_b&=&
\frac{e_K|{\bm k}_\pi|}{3}\frac{D_{\bar{K}p}D_{\pi\Sigma}}
{E_{\gamma}+E_p-E_K-M_{\Sigma^*}},
\nonumber\\
V_c&=&\frac{e_\pi|{\bm k}_K|}{3}
\frac{D_{\bar{K}p}D_{\pi\Sigma}}{E_p-E_K-M_{\Sigma^*}},
\end{eqnarray}
%EQUATION<<<
where $e_\pi$ and $e_K$ denote the unit electric charges for the outgoing pion and kaon, respectively. Similarly to the manipulation in the previous section, we can write the total scattering amplitude as follows:
%EQUATION>>>
\begin{eqnarray}
\label{eq:Tg}
T_{\gamma}&=&\underbrace{
[i{\bm\sigma}\cdot(\hat{\bm\epsilon}\times\hat{\bm k}_\gamma)]
T_a}_\mathrm{(a')}
\nonumber\\
&+&\underbrace{[2(\hat{\bm k}_\pi\cdot\hat{\bm\epsilon})
-i{\bm\sigma}\cdot(\hat{\bm k}_\pi\times\hat{\bm\epsilon})]
T_b}_\mathrm{(b')}
+\underbrace{[2(\hat{\bm\epsilon}\cdot\hat{\bm k}_K)
-i{\bm\sigma}\cdot(\hat{\bm\epsilon}\times\hat{\bm k}_K)]
T_c}_\mathrm{(c')},
\end{eqnarray}
%EQUATION<<<
where
%EQUATION>>>
\begin{eqnarray}
\label{eq:ETg}
T_a&=&V_a+V_a\,G\,T_\mathrm{WT},
\nonumber\\
T_b&=&V_b+V_b\,G\,T_{\Sigma^*},
\nonumber\\
T_c&=&V_c+V_c\,G\,T_\mathrm{WT}.
\end{eqnarray}
%EQUATION<<<
In the numerical calculations, as for the meson-baryon scattering amplitudes, $T_\mathrm{WT}$ and $T_{\Sigma^{*}}$ in Eq.~(\ref{eq:WTSIG}), and the meson-baryon propagator, $G$ in Eq.~(\ref{eq:GDIM}), we have substituted $\sqrt{s}$ with the $\pi\Sigma$ invariant mass, $M_{\pi\Sigma}$.

%--------------------------------------------------
\section{Numerical results}
%--------------------------------------------------
In this section, we present the numerical results for the $\gamma p\to K^+ \pi^\pm \Sigma^\mp$ reaction process, using the $\chi$UM. First, we show the $\pi\Sigma$ invariant-mass spectrum as a function of $M_{\pi\Sigma}$ at $E_\gamma=1.7$ GeV in Fig.~\ref{fig2}, using Eq.~(\ref{eq:TDC}) defined in Appendix.  We depict the two different isospin channels, $\pi^+\Sigma^-$ (left  panel) and $\pi^-\Sigma^+$ (right panel), separately. The thin lines denote the case without the $\Sigma^*$-pole contribution, whereas thick ones with it. As shown in the figure, the effect of the $\Sigma^*$-pole contribution plays an important role to produce the shape of the invariant-mass spectrum. However, the interference pattern between the two isopsin channels are different in such way that it is destructive for the $\pi^+\Sigma^-$ channel, whereas constructive for the $\pi^-\Sigma^+$ one. This behavior can be understood approximately by the fact that there is a sign difference in the coefficient $C^*$ in Eq.~(\ref{eq:DD}) for the two channels: $C^*_{\pi^+\Sigma^-}=-1/\sqrt{2}$ and $C^*_{\pi^-\Sigma^+}=1/\sqrt{2}$ for the $\Sigma^*$-pole contribution. 

The strength of the invariant-mass spectrum turns out to be qualitatively larger for the $\pi^-\Sigma^+$ channel than that for the $\pi^+\Sigma^-$ one. This tendency is rather different from that observed in the recent LEPS experiment~\cite{Niiyama:2008rt}: $\pi^-\Sigma^+<\pi^+\Sigma^-$. Although we do not show the results for higher photon energy being larger than $1.7$ GeV, we verified that the strength of the invariant-mass spectrum increases with respect $E_\gamma$ for both isospin channels. Again, this energy dependence is not consistent with the experimental observation as mentioned in Section~I. This unfavorable results in the present approach, in comparison to the experiment, may tells us that more complicated and realistic contributions, which reflect the internal structure of $\Lambda^*$ for instance, are necessary. 

Finally, in Fig.~\ref{fig3}, we show the Dalitz plots as a function of $M_{\pi\Sigma}$ and $M_{K\pi}$ for the two different isospin channels, $\pi^+\Sigma^-$ (left panel) and $\pi^-\Sigma^+$ (right panel), at $E_\gamma=1.7$ GeV. Note that, since the kinematically possible region for the Dalitz plot at $E_\gamma=1.7$ GeV contains only negligible effects from $K^{*0}(892)$, which decays into $\pi^-{K}^+$, we have not considered the $K^{*0}(892)$-pole contribution in the present work. As the photon energy grows, it is necessary for us to consider its effects, making an interference pattern between the $K^*$ and $(\Lambda^*+\Sigma^*)$ contributions. We note that this sort of interferences may play an important role to understand complicated aspects of resonance findings. It is worth mentioning that a related argument was done for a method to find the exotic pentaquark $\Theta^+$ with more larger statistics in Ref.~\cite{Amarian:2006xt}.  

As seen in the Dalitz plots, the peak structure is rather different between the two isospin channels; the effect from the $\Sigma^*$-pole contribution becomes weak  as $M_{\pi K}$ increases for the $\pi^-\Sigma^+$ channel, whereas it remains almost the same for the $\pi^+\Sigma^-$ one. This tendency is caused by a complicated interference between the resonant contributions.

%FIGURE>>>
\begin{figure}[t]
\begin{tabular}{cc}
\includegraphics[width=8cm]{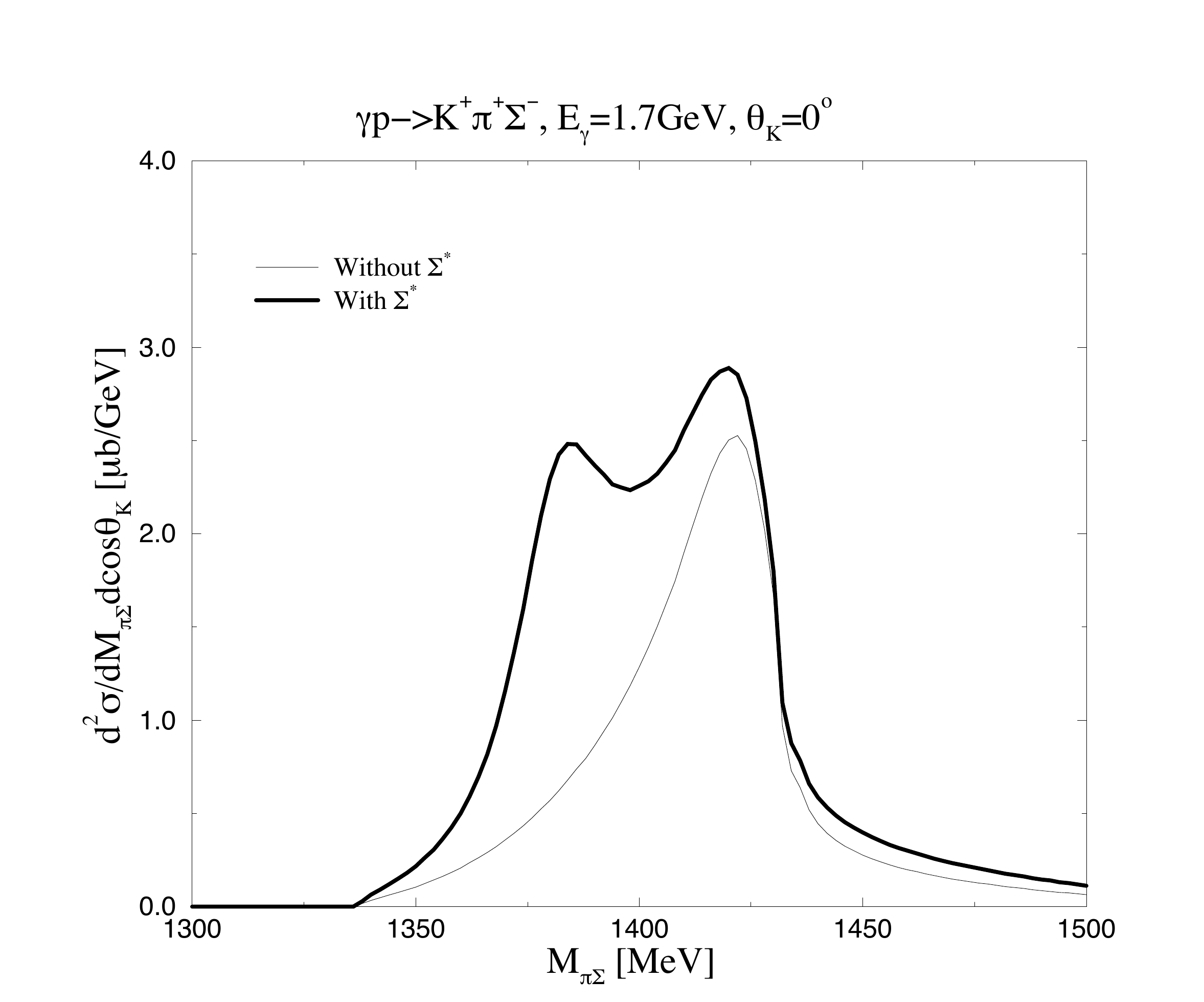}
\includegraphics[width=8cm]{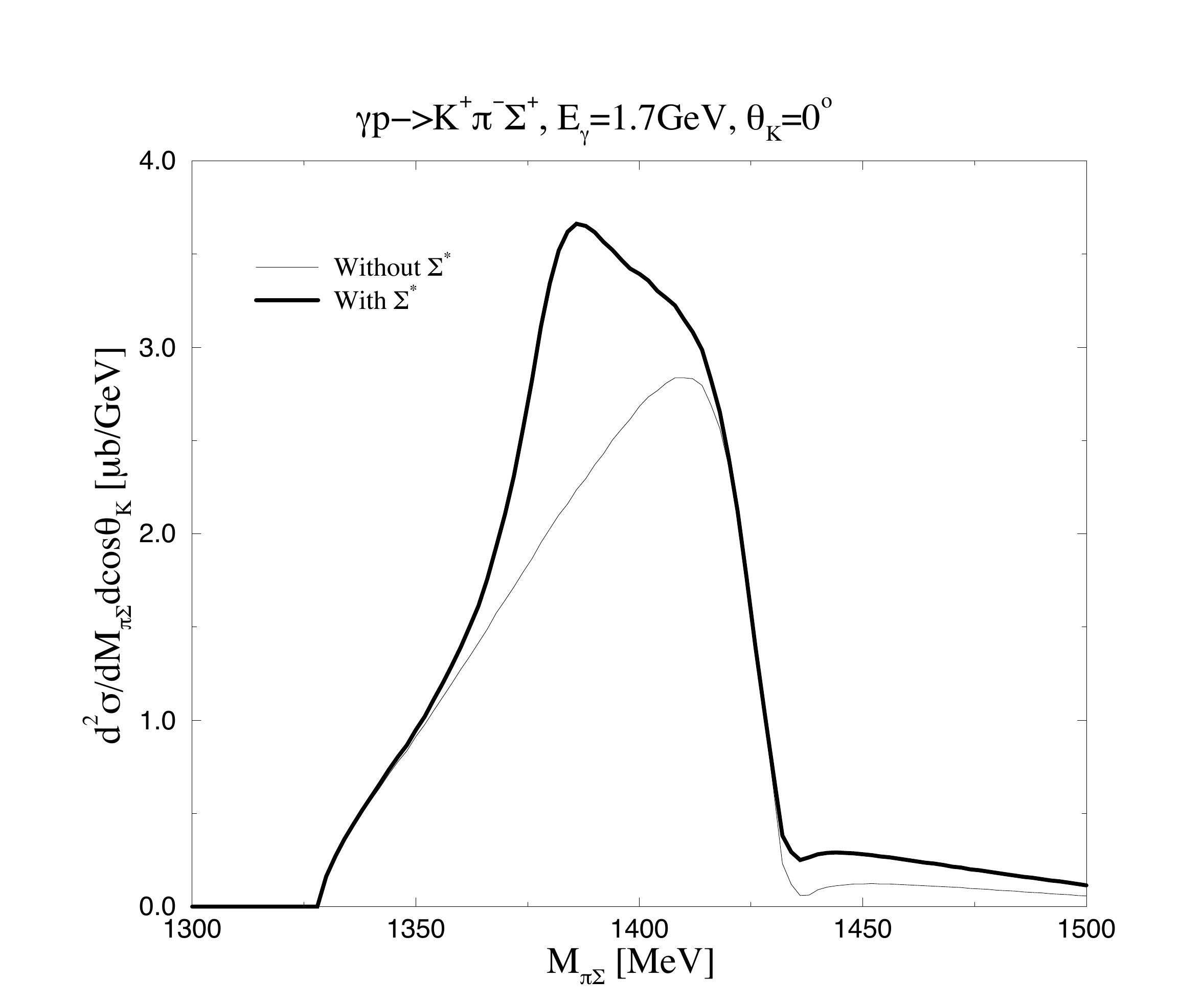}
\end{tabular}
\caption{$\pi\Sigma$ invariant-mass spcetrum as a function of $M_{\pi\Sigma}$ for the two different isospin channels, $\pi^+\Sigma^-$ (left) and $\pi^-\Sigma^+$ (right), at $E_\gamma=1.7$ GeV. Thin lines denote the case without the $\Sigma^*$ contribution, whereas thick ones with it.}        
\label{fig2}
\end{figure}
%FIGURE<<< 

%FIGURE>>>
\begin{figure}[t]
\begin{tabular}{cc}
\includegraphics[width=8cm]{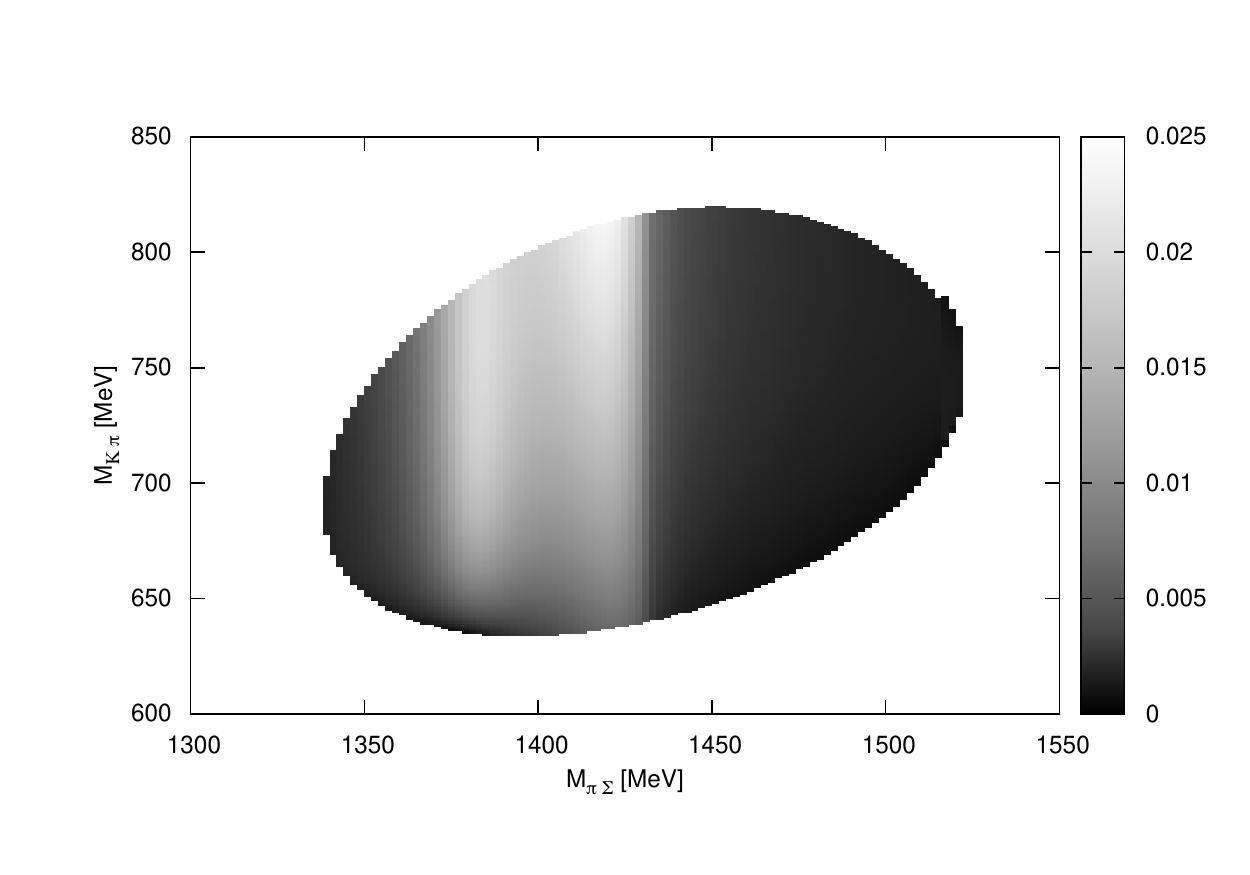}
\includegraphics[width=8cm]{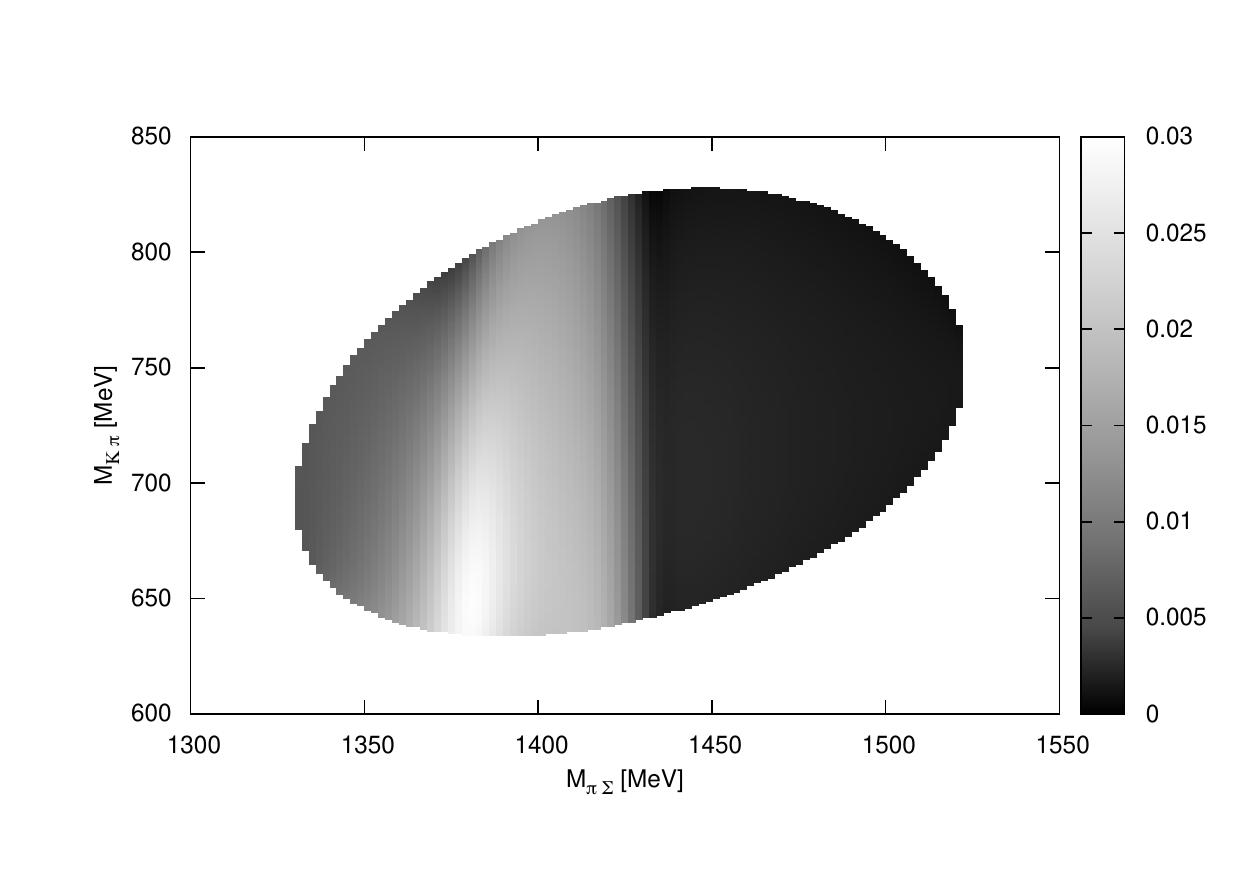}
\end{tabular}
\caption{Dalitz plot as a function of $M_{\pi\Sigma}$ and $M_{K\pi}$ for the two different isospin channels, $\pi^+\Sigma^-$ (left) and $\pi^-\Sigma^+$ (right), at $E_\gamma=1.7$ GeV.}        
\label{fig3}
\end{figure}
%FIGURE<<< 

%FIGURE>>>
\begin{figure}[t]
\includegraphics[width=10cm]{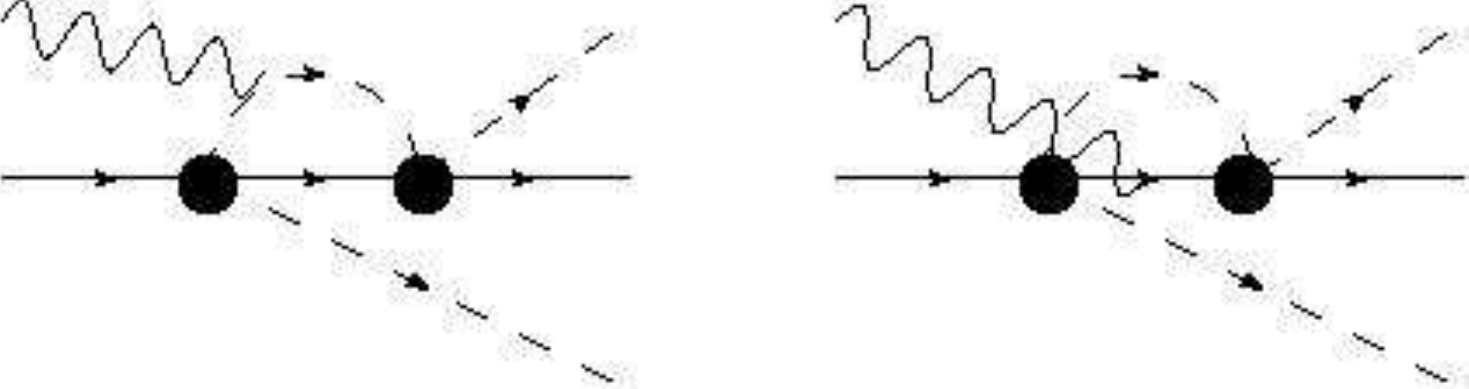}
\caption{Feynman diagrams with the meson-baryon loop, to which the incident photon couples.}        
\label{fig4}
\end{figure}
%FIGURE<<< 

%--------------------------------------------------
\section{Summary and conclusion}
%--------------------------------------------------
We have studied the $\Lambda(1405,1/2^-)\equiv\Lambda^*$ photoproduction, using the chiral unitary model, with the $p$-wave $\Sigma^*$-pole contribution. The scattering amplitude was computed by unitarizing the element $\gamma p\to K^+\pi^\pm\Sigma^\mp$ scattering process, which was derived from the Weinberg-Tomozawa meson-baryon interaction with the $\Sigma^*$-pole contribution. All the necessary parameters, such as the renormalization scale, were determined to reproduce the experimental data for the $S=-1$ meson-baryon scattering processes with $p$-wave contributions. 

As for the numerical results, we presented the $\pi\Sigma$ invariant-mass spectrums as well as the Dalitz plots for the two different isospin channels, $\pi^\pm\Sigma^\mp$. We observed that the $\Sigma^*$-pole contribution plays an important role to produce the invariant-mass spectrum. In contrast to  the recent LEPS experiment, the strength of the invariant-mass spectrum was slightly larger for the $\pi^-\Sigma^+$ channel than the $\pi^+\Sigma^-$ one. Moreover, the experimentally observed energy dependence of the $\Lambda^*$-production rate was not seen in the present results. The analysis on the Dalitz plot showed that there were obvious different interference patterns for the resonance contributions for the two ispspin channels.   

Considering these unfavorable results in comparison to the experimental data, it seems necessary for us to incorporate more realistic and complicated contributions, such as the meson-baryon loop, to which  the incident photon couples, etc., as shown in Fig.~\ref{fig4}. From this sort of contributions, one may extract information on the internal structure of $\Lambda^*$, which has been suggested to be different from usual color-singlet baryons. Moreover, since $p$-wave contributions can appear even from the Weinberg-Tomozawa interaction, if we go beyond the leading  contributions, there can be some extend of modifications for the present results. It is worth mentioning that another new experiment for the $\Lambda^*$-photoproduction has been performed and under analyses by LEPS collaboration at SPring-8~\cite{Ahn}.  More detailed theoretical works are underway and will appear elsewhere.

%-------------------------------------------------
\section*{acknowledgment}
%-------------------------------------------------
The authors are grateful especially to the organizers for the workshop {\it Strangeness in multi-quark systems}, which was held during $29-31$ October 2008 in Kaga, Japan (http://nexus.kek.jp/Tokutei/workshop/2008/index.htm). The works of S.i.N. and D.J. were partially supported by the Grant for Scientific Research (Priority Area No.17070002, No.20028005, and No.20028004) from the Ministry of Education, Culture, Science and Technology (MEXT) of Japan. This work was also done under the Yukawa International Program for Quark-Hadron Sciences. The numerical calculations were carried out on YISUN at YITP in Kyoto University. 
%-------------------------------------------------
\section*{Appendix}
%-------------------------------------------------
Considering a specific experimental setup, such as that of LEPS collaboration at SPring-8, we are interested in the case that the outgoing $K^+$ is detected in a very forward angle, $\theta_K\approx0$, in the center of mass frame. This condition simplifies the kinematics of the present scattering process considerably, and we can define the following double-differential cross section as a function of the $\pi\Sigma$ invariant mass $M_{\pi\Sigma}$ and the angle $\theta_K$:   
%EQUATION>>>
\begin{equation}
\label{eq:TDC}
\frac{d^2\sigma}{dM_{\pi\Sigma}\,d\cos\theta_K}
=\int\frac{M_{K\pi}\,M_{\pi\Sigma}\,dM_{K\pi}}
{128\,\pi^3\,s^{3/2}\,E_\gamma}\,|\overline{\mathcal{M}}|^2,
\end{equation}
%EQUATION<<<
where the outgoing $\Sigma$ is scattered isotrophically for the azimuthal angle $\phi_{\Sigma}$ as a consequence of $\theta_K\approx0$. According to the specific kinematical condition, the four momenta of the relevant particles can be written in the center of mass frame as follows:
%EQUATION>>>
\begin{eqnarray}
\label{eq:momen}
k_\gamma&=&(E_\gamma,0,0,k),
\nonumber\\
k_p&=&(E_p,0,0,-k),
\nonumber\\
k_K&=&(E_K,0,0,p_K),
\nonumber\\
k_\pi&=&(\sqrt{s}-E_K-E_\Sigma,0,-p_\Sigma\sin\theta_\Sigma,
-p_K-p_\Sigma\cos\theta_\Sigma),
\nonumber\\
k_\Sigma&=&(E_\Sigma,0,p_\Sigma\sin\theta_\Sigma,
p_\Sigma\cos\theta_\Sigma).
\end{eqnarray}
%EQUATION<<<
Here, $E_h$ stands for the relativistic energy for a particle $h$, $E=\sqrt{M^2_h+|{\bm k_h}|^2}$. The absolute values of the three momenta for $K^+$ and $\Sigma$, $p_K$ and $p_\Sigma$, respectively, become
%EQUATION>>>
\begin{equation}
\label{eq:}
p_K=\left[\left(\frac{s+M^2_K-M^2_{\pi\Sigma}}{2\sqrt{s}}\right)^2
-M^2_K\right]^{1/2},\,\,\,\,
p_\Sigma=\left[\left(\frac{s+M^2_\Sigma-M^2_{K\pi}}{2\sqrt{s}}\right)^2
-M^2_\Sigma\right]^{1/2}.
\end{equation}
%EQUATION<<<
The angle $\theta_\Sigma$ is determined from the following condition that
%EQUATION>>>
\begin{equation}
\label{eq:the}
\cos\theta_\Sigma
=\frac{(\sqrt{s}-E_K-E_\Sigma)^2-M^2_\pi-p^2_K-p^2_\Sigma}
{2p_Kp_\Sigma},
\end{equation}
%EQUATION<<<
which defines the kinematically allowed region, confined by the minimum and maximum values of the Dalitz plot:
%EQUATION>>>
\begin{equation}
\label{eq:DalC}
(M_K+M_\pi)^2\le M^2_{K\pi}\le(\sqrt{s}-M_\Sigma)^2,
\,\,\,\,
(M_\pi+M_\Sigma)^2\le M^2_{\pi\Sigma}\le(\sqrt{s}-M_K)^2.
\end{equation}
%EQUATION<<<
$|\overline{\mathcal{M}}|^2$ indicates the squared invariant amplitude, averaged over the spin and polarization:
%EQUATION>>>
\begin{equation}
\label{eq:AA}
|\overline{\mathcal{M}}|^2=\frac{1}{4}\sum_\mathrm{spin,pol}|\mathcal{M}|^2.
\end{equation}
%EQUATION<<<

%--------------------------------------------------

\end{document}